\documentclass[a4paper,usenatbib,fleqn]{mnras}

\usepackage[T1]{fontenc}
\usepackage{ae,aecompl}

\usepackage{graphicx}	
\usepackage{amsmath}	
\usepackage{amssymb}	

\title[Intermittency of Dissipation in Alfv\'{e}nic Turbulence]{Intermittency of Energy Dissipation in Alfv\'{e}nic Turbulence}

\author[V. Zhdankin et al.]{
Vladimir Zhdankin,$^{1}$\thanks{E-mail: zhdankin@jila.colorado.edu}
Stanislav Boldyrev,$^{2,3}$
Christopher H.K. Chen$^{4}$
\\
$^{1}$JILA, NIST and University of Colorado, 440 UCB, Boulder, Colorado 80309, USA\\
$^{2}$Department of Physics, University of Wisconsin-Madison, 1150 University Avenue, Madison, Wisconsin 53706, USA\\
$^{3}$Space Science Institute, Boulder, Colorado 80301, USA\\
$^{4}$Department of Physics, Imperial College London, London SW7 2AZ, UK
}

\date{Accepted XXX. Received YYY; in original form ZZZ}

\pubyear{2015}

\begin{document}
\label{firstpage}
\pagerange{\pageref{firstpage}--\pageref{lastpage}}
\maketitle

\begin{abstract}
We investigate the intermittency of energy dissipation in Alfv\'{e}nic turbulence by considering the statistics of the coarse-grained energy dissipation rate, using direct measurements from numerical simulations of magnetohydrodynamic turbulence and surrogate measurements from the solar wind. We compare the results to the predictions of the log-normal and log-Poisson random cascade models. We find that, to a very good approximation, the log-normal model describes the probability density function for the energy dissipation over a broad range of scales, but does not accurately describe the scaling exponents of the moments. The log-Poisson model better describes the scaling exponents of the moments, while the comparison with the probability density function is not straightforward.
\end{abstract}

\begin{keywords}
turbulence -- plasmas -- MHD -- solar wind
\end{keywords}



\section{Introduction}

Intermittent energy dissipation, caused by the stochastic disposition of the turbulent energy cascade as it proceeds from large scales to small scales, is a hallmark of turbulence. It leads to dynamics that are inhomogeneous in space and in time, as manifest by coherent structures and intense dissipative events. For turbulent plasmas, which are often modeled at large scales by magnetohydrodynamics (MHD), intermittency forms sites for magnetic reconnection, particle heating, and particle acceleration. It therefore plays a central role in many space and astrophysical systems, governing, for example, the heating of the solar corona and the solar wind \citep{cranmer_etal_2007, osman_etal_2011, uritsky_etal_2007}, granulation in the solar photosphere \citep{cattaneo_1999, stein_nordlund_2006}, star formation in the interstellar medium \citep{pan_etal_2009}, and flares from accreting systems \citep{dimatteo_etal_1999, albert_etal_2007} and pulsar wind nebulae \citep{tavani_etal_2011,abdo_etal_2011}. 

Intermittency remains one of the most challenging aspects of turbulence to describe theoretically. Nevertheless, an attractive phenomenological picture exists in random cascade models. The earliest of these was the log-normal model, developed by Kolmogorov to describe the stochastic fragmentation of the energy cascade in hydrodynamic turbulence \citep{kolmogorov_1962}. In this model, the coarse-grained energy dissipation rate $\epsilon_l$, defined as the energy dissipation rate averaged across regions of linear size $l$, approaches a log-normal distribution at small scales. The variance of the probability density function (PDF) $P(\epsilon_l)$ increases as $l$ decreases, giving the dynamics an explicit scale dependence. The log-normal model was followed by a number of other random cascade models, culminating in the log-Poisson model \citep{she_waymire_1995, dubrulle_1994}. When Kolmogorov's refined similarity hypothesis is assumed, these models predict the scaling exponents of velocity structure functions.

Random cascade models successfully describe numerical and experimental observations of intermittency in hydrodynamic turbulence, although it can be difficult to differentiate the models. The log-normal model is often consistent with observations \citep{naert_etal_1998, arneodo_etal_1998, arneodo_etal_1999, molchan_1997, almalkie_debruynkops_2012}, despite the fact that it is now known to be inconsistent for incompressible hydrodynamic turbulence, since it implies a non-monotonic scaling of velocity structure function exponents when the refined similarity hypothesis is assumed \citep[see, e.g.,][]{frisch1995}. The scaling exponents of higher-order structure functions are generally better fit by the log-Poisson model \citep{she_leveque_1994}.

In contrast to hydrodynamic turbulence, a convincing phenomenological picture of intermittency in a plasma, where the inertial-range turbulent cascade is mediated by Alfv\'{e}n waves \citep{goldreich1995}, remains to be discovered. A number of studies inferred intermittency from the scaling exponents of structure functions, both in numerical simulations \cite[e.g.,][]{politano_pouquet_1998b, muller_biskamp_2000, muller_etal_2003, cho_etal_2002} and in the solar wind \cite[e.g.,][]{burlaga_1993, horbury_balogh_1997, chen_etal_2014}. Several models based on log-Poisson statistics were developed to explain these observations \citep{grauer_etal_1994, politano_pouquet_1995, muller_biskamp_2000, chandran_etal_2014}. In the present paper, we pursue a more direct route to measuring intermittency, based on the statistics of energy dissipation rather than field fluctuations. The energy dissipation rate has a clear interpretation in random cascade models, being tied to the energy cascade rate, and does not require the refined similarity hypothesis to be assumed. The intermittency of energy dissipation is evident from the formation of current sheets in MHD turbulence \cite[e.g.,][]{zhdankin_etal_2014}, but only a handful of previous studies carefully investigated the statistics of coarse-grained dissipation \citep{biskamp_1995, merrifield_etal_2005, bershadskii_2003}.

In this Letter, we investigate the intermittency of the Alfv\'{e}nic energy cascade by considering the statistics of the coarse-grained energy dissipation rate in numerical simulations and in the solar wind, using a surrogate quantity in the latter case. We show that, in both systems, the PDFs are robustly consistent with the log-normal model, while the scaling of higher-order moments can be better described by the log-Poisson model.

\section{Results}

We analyze MHD turbulence simulations and inertial-range solar wind measurements. The incompressible MHD equations for plasma velocity $\boldsymbol{v}(\boldsymbol{x},t)$ and magnetic field $\boldsymbol{B}(\boldsymbol{x},t)=\boldsymbol{B}_0+\boldsymbol{b}(\boldsymbol{x},t)$ (where $\boldsymbol{B}_0 = B_0 \hat{\boldsymbol{z}}$ is the uniform background field) are
\begin{eqnarray}
\partial_t \boldsymbol{v} + (\boldsymbol{v} \cdot \nabla) \boldsymbol{v} &=& - \nabla p + (\nabla \times \boldsymbol{B}) \times \boldsymbol{B} + \nu \nabla^2 \boldsymbol{v} + \boldsymbol{f}_1, \nonumber \\
\partial_t \boldsymbol{B} &=& \nabla \times (\boldsymbol{v} \times \boldsymbol{B}) + \eta \nabla^2 \boldsymbol{B} +\boldsymbol{f}_2, \label{eq:mhd}
\end{eqnarray}
along with $\nabla \cdot \boldsymbol{v} = \nabla \cdot \boldsymbol{B} = 0$. In the limit of $B_0 \gg b$ and for wavevectors $k_\perp \gg k_\parallel$, reduced MHD (RMHD) is obtained from Eqs.~\ref{eq:mhd} by retaining only the components of $\boldsymbol{b}$ perpendicular to $\boldsymbol{B}_0$. The resistive and viscous energy dissipation rates (per unit volume) are given by $\epsilon^\eta = \eta j^2 = \eta |\nabla \times \boldsymbol{B}|^2$ and $\epsilon^\nu = 2 \nu \sigma_{ij} \sigma_{ij}$, where $\eta$ the resistivity, $\nu$ is the viscosity, and $\sigma_{ij} = \frac{1}{2} \left( \frac{\partial v_j}{\partial x_i} + \frac{\partial v_i}{\partial x_j} \right)$ is the rate-of-strain tensor.

We consider numerical simulations from two sets of codes: one solves the full MHD equations while the other solves the RMHD equations \citep{cattaneo_etal_2003, perez_etal2012}. Both codes solve the respective equations in a periodic box of size $(L_\perp, L_\perp, L_\parallel)$ using standard pseudospectral methods. For $B_0/b_\text{rms} \le 1$, $L_\parallel = L_\perp$; for $B_0/b_\text{rms} > 1$, $L_\parallel$ is elongated by a similar factor to accommodate the anisotropy. Turbulence is driven at the largest scales by applying random forces $\boldsymbol{f}_1$ and $\boldsymbol{f}_2$ in Fourier space at wave-numbers $2\pi/L_\perp \leq k_{x,y} \leq 2 (2\pi/L_\perp)$, $k_z = 2\pi/L_\parallel$. The correlation between the forces is chosen to resemble independent counter-propagating shear-Alfv\'en modes. The RMHD simulations have Reynolds numbers $\operatorname{Re} = v_{\rm rms} (L_\perp/2\pi) / \nu$ ranging from $1000$ to $9000$ (on grids ranging from $512^3$ to $2048^3$), while the full MHD simulations have $\operatorname{Re} \approx 2200$ (on a $512^3$ grid). We set $\nu = \eta$.

We also consider measurements from the solar wind taken by the Ulysses spacecraft \citep{wenzel_etal_1992}. We use 1s resolution magnetic field data \citep{balogh_etal_1992} during a 30 day interval (days 100-129 of 1995) when Ulysses was in a steady fast wind stream at 1.4 AU from the Sun containing large amplitude ($\delta B/B\sim 1$) Alfv\'{e}nic fluctuations at large scales \citep{chen_etal_2012}. The temporal data from the solar wind is assumed to be equivalent to spatial data by the Taylor hypothesis.

Following the notation in \cite{biskamp2003}, we consider line-averaged energy dissipation rates at scales $l_n = 2^{-n} L$, where $L$ is the box size in simulations and the size of the overall data interval in the solar wind. The coarse-grained resistive and viscous energy dissipation rates at point $\boldsymbol{r}$, averaged at the $n$th level, are respectively given by 
\begin{align}
\epsilon^\eta_n(\boldsymbol{r}) &= \frac{1}{l_n} \int_{-l_n/2}^{l_n/2} d\xi \epsilon^\eta (\boldsymbol{r} + \xi \hat{\boldsymbol{x}}) , \nonumber \\
\epsilon^\nu_n(\boldsymbol{r}) &= \frac{1}{l_n} \int_{-l_n/2}^{l_n/2} d\xi \epsilon^\nu (\boldsymbol{r} + \xi \hat{\boldsymbol{x}}) \, .
\end{align}
where $\hat{\boldsymbol{x}}$ is the direction along the measurement interval, taken perpendicular to the guide field in simulations. The coarse-grained energy dissipation rate is then $\epsilon_{l_n} \equiv \epsilon_n = \epsilon^\eta_n + \epsilon^\nu_n$. We considered other shapes for the averaging regions (i.e., cubes and squares) in simulations, and found our conclusions to be insensitive to this shape; results are mainly from line averages to ease comparison with the solar wind.

\begin{figure*}
\includegraphics[width=8cm]{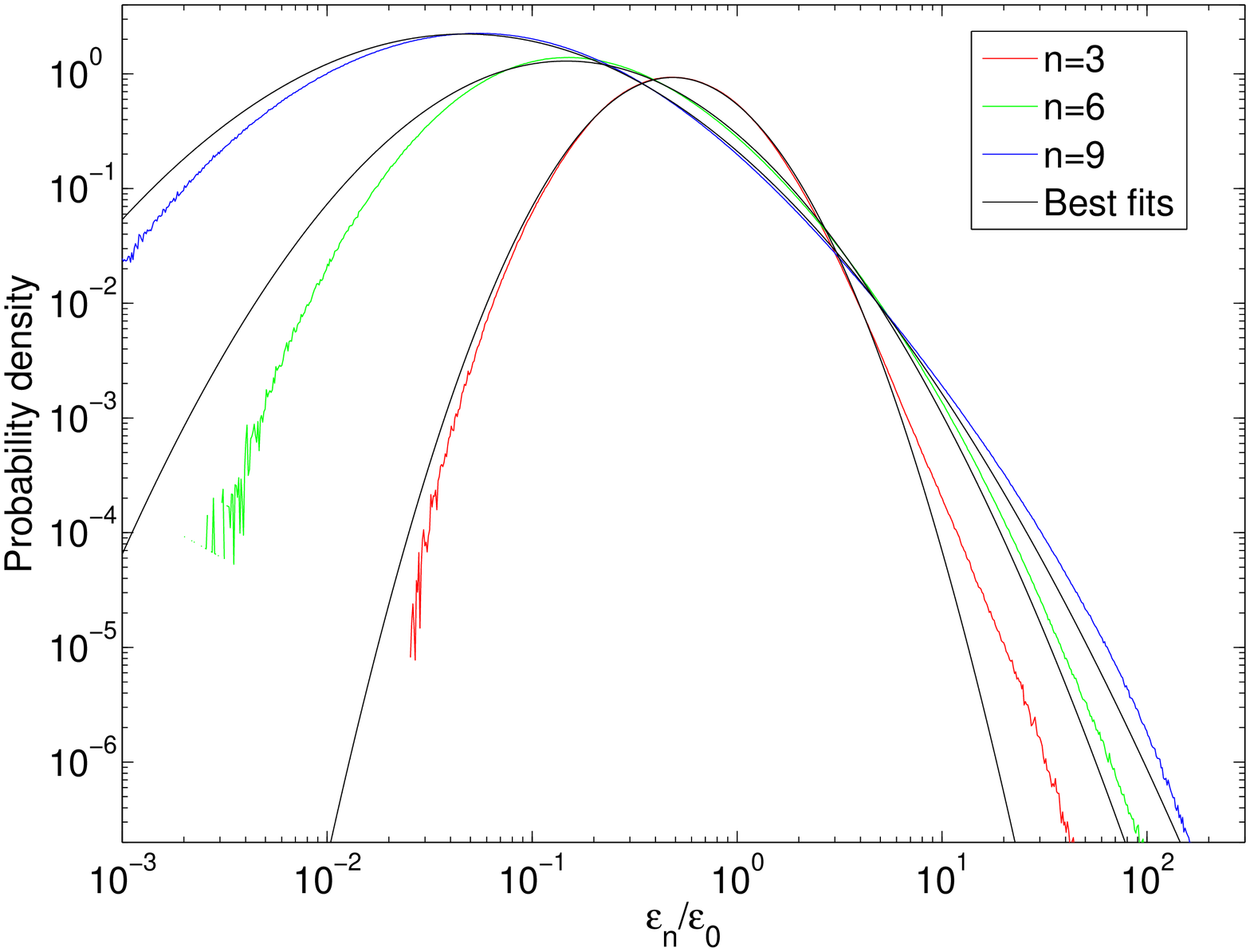}
\includegraphics[width=8cm]{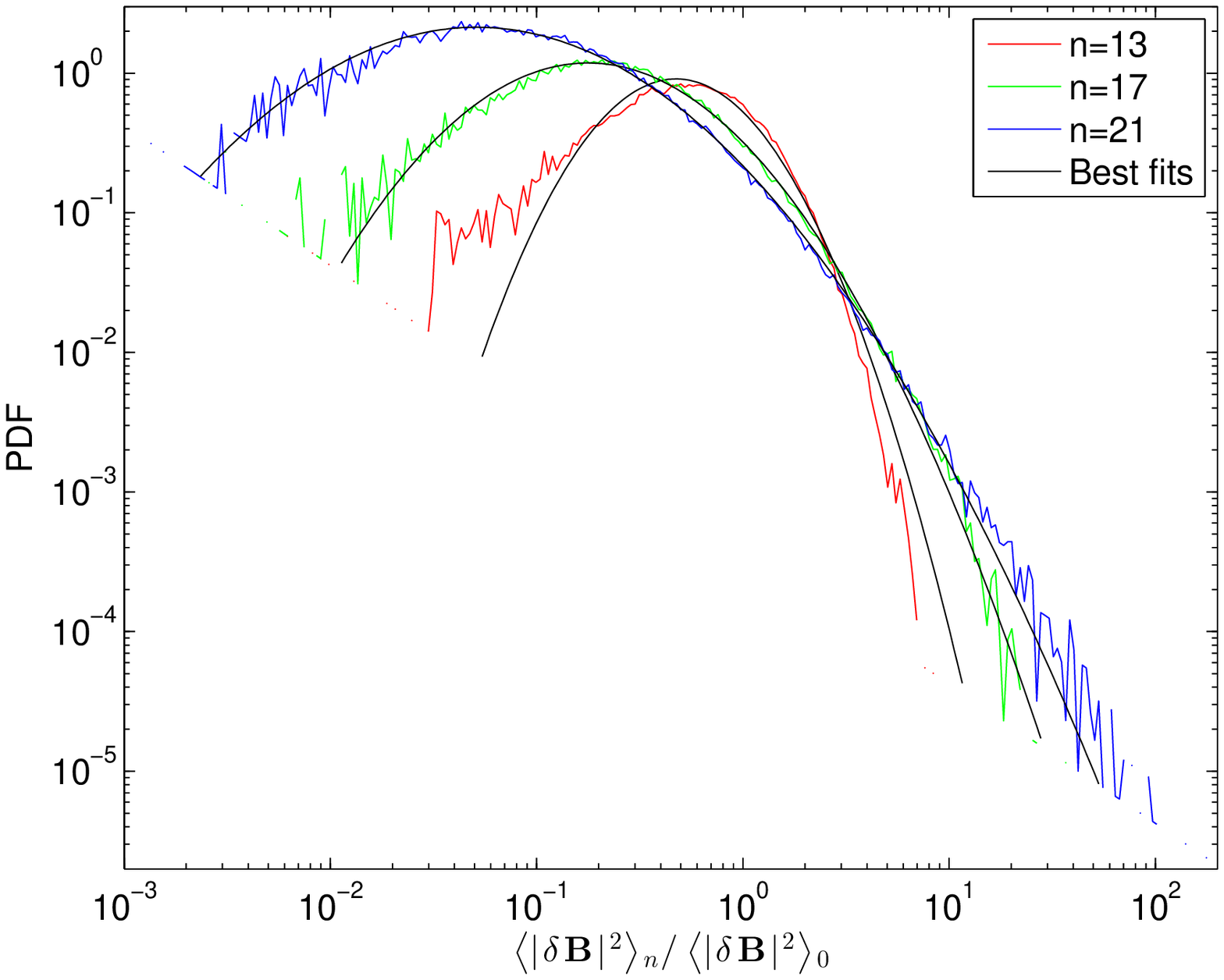}
   \centering
   \caption{\label{fig:dist_energy} Left panel: The PDF for coarse-grained energy dissipation rate, $\epsilon_n/\epsilon_0$, in RMHD simulations for $n \in \{3,6,9\}$ (in red, green, and blue, respectively). Right panel: The PDF for the surrogate dissipation, $\chi_n/\chi_0 = \langle |\delta \boldsymbol{B}|^2 \rangle_n / \langle |\delta \boldsymbol{B} |^2\rangle_0$, in solar wind data for $n \in \{13,17,21\}$ (in red, green, and blue, respectively). Best-fit log-normals are shown in black.}
 \end{figure*}

Representative measurements of the PDF $P(\epsilon_n/\epsilon_0)$ from a simulation are shown in the left panel of Fig.~\ref{fig:dist_energy} for RMHD with $\operatorname{Re} = 9000$. We find that $P(\epsilon_n/\epsilon_0)$ is fit remarkably well by a log-normal distribution, 
\begin{eqnarray}
P(\epsilon_n/\epsilon_0) = \frac{1}{\sqrt{2\pi \sigma_n^2}} \frac{\epsilon_0}{\epsilon_n} \exp{\left[-\frac{1}{2\sigma_n^2}\left(\log{\frac{\epsilon_n}{\epsilon_0}} - \mu_n \right)^2 \right]},\quad
\end{eqnarray}
where $\mu_n$ and $\sigma_n$ are the location parameter and scale parameter, respectively. The log-normal distribution provides a reasonable fit to the bulk of $P(\epsilon_n/\epsilon_0)$ for all scales in all of the given simulations, with typical relative errors of a few percent. For example, at moderately large $n$, local relative errors are less than $5\%$ between $\epsilon_n/\epsilon_0 \approx 10^{-2}$ to $\epsilon_n/\epsilon_0 \approx 10$. At small to intermediate $n$, the PDF systematically exceeds the fit at large values ($\epsilon_n/\epsilon_0 \gtrsim 10$), while it is below the fit at small values of $\epsilon_n/\epsilon_0$.

Since direct measurements of the energy dissipation are unavailable for the solar wind, we use $|\delta{\boldsymbol{B}}|^2(\boldsymbol{x}) = |\boldsymbol{B}(\boldsymbol{x}+\delta{\boldsymbol{x}}) - \boldsymbol{B}(\boldsymbol{x})|^2$ as a surrogate for $j^2(\boldsymbol{x})$, where $\delta{\boldsymbol{x}}$ is some small increment. Although the dissipation in the solar wind is due to collisionless mechanisms of dissipation rather than viscous and resistive dissipation, we will assume $j^2$ to be a reasonable proxy, based on correlations in numerical simulations of collisionless plasmas \citep[e.g.,][]{tenbarge_etal_2013, wan_etal_2015, makwana_etal_2015}. We denote the average of $|\delta{\boldsymbol{B}}|^2$ across the interval of length $l_n$ as $\chi_n = \langle |\delta{\boldsymbol{B}}|^2\rangle_n$, which is analogous to $\epsilon^\eta_n$. In our simulations, we find that $P(\chi_n/\chi_0)$ is well fit by a log-normal for $l_n > \delta x$, and that best agreement with $P(\epsilon_n/\epsilon_0)$ is obtained for scales below $\delta{x} \approx b_\text{rms}/j_\text{rms}$, which is roughly the scale at which resistive dissipation is maximized. Accordingly, for the solar wind, we compute $|\delta{\boldsymbol{B}}|^2$ using $\delta x$ at the ion gyroscale. Representative measurements of $P(\chi_n/\chi_0)$ for the solar wind are shown in the right panel of Fig.~\ref{fig:dist_energy}. Log-normal fits work well when $4 \le n \le 21$, where $n = 0$ corresponds to the entire solar wind data interval and $n = 21$ corresponds to the ion gyroscale.

We now consider the scaling properties of the log-normal fits. Since the variables are normalized to the mean, the parameters satisfy $\mu_n = -\sigma_n^2/2$, leaving one free parameter for each $n$. Kolmogorov's log-normal model predicts that $\sigma_n^2/2 \propto n$, with the coefficient of proportionality given by $(\ln{2}/2) m$, where the intermittency parameter $m > 0$ describes the degree of intermittency (with $m = 0$ for non-intermittent dynamics) \citep[e.g.,][]{biskamp2003, sreenivasan_1993}.

\begin{figure}
\resizebox{\columnwidth}{!}{\includegraphics{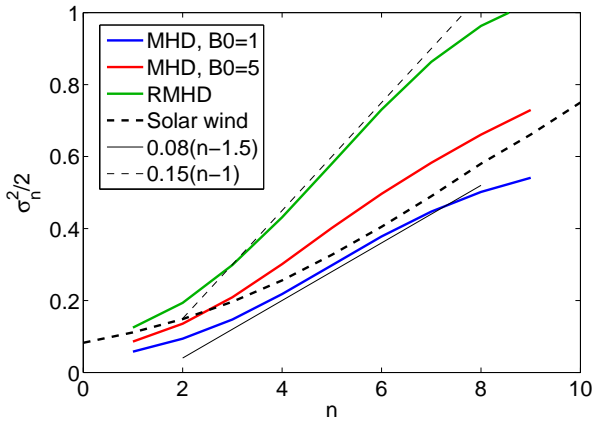}}
\caption{\label{fig:variance_updated} The scaling of $\sigma_n^2/2$ versus subdivision level $n$, obtained from best-fit log-normals to the PDFs for coarse-grained energy dissipation rates. Results are shown for total dissipation rate in MHD simulations with $B_0 \lesssim 1$ (blue) and $B_0 \approx 5$ (red), total dissipation rate in RMHD simulations (green), and surrogate dissipation rate in the solar wind (thick black dashed line). The scalings vary from $\sigma_n^2/2 \propto 0.08n$ for weak guide fields to $\sigma_n^2/2 \propto 0.15n$ for strong guide fields.}
\end{figure}

The scaling of $\sigma_n^2/2$, obtained from the best-fit log-normal distributions, is shown in Fig.~\ref{fig:variance_updated} for several cases: MHD with weak guide field ($B_0 \lesssim b_\text{rms}$), MHD with strong guide field ($B_0 \approx 5 b_\text{rms}$), RMHD, and the solar wind. We identify a range of intermediate scales where $\sigma_n^2/2 \propto n$, with slopes implying an intermittency parameter that ranges from $m \approx 0.23$ for weak guide fields (as in the hydrodynamic case) to $m \approx 0.43$ for RMHD. This result agrees with previous numerical studies showing that intermittency increases for stronger guide fields \citep{muller_etal_2003}. The intermittency in full MHD approaches that of RMHD when the guide field is increased, but does so quite slowly - there is a significant difference even when $B_0/b_\text{rms} \approx 5$. Thus, RMHD may be inaccurate for describing intermittency in these cases, likely due to the condition $\delta b / B_0 \ll 1$ breaking down locally in MHD turbulence for the given guide fields. The linear scaling of $\sigma_n^2/2$ with $n$ is insensitive to $\operatorname{Re}$ and the shape of the averaging regions, although these inputs may shift the curve. We note that $\sigma_n^2/2$ has a somewhat steeper scaling for the resistive and viscous contributions alone, i.e., $\epsilon^\eta_n/\epsilon^\eta_0$ and $\epsilon^\nu_n/\epsilon^\nu_0$, evidently due to the anti-correlation of $j^2$ and $\omega^2$ in the dissipation range.

For comparison, the scaling of $\sigma_n^2/2$ from log-normal fits to the PDF $P(\chi_n/\chi_0)$ for the surrogate variable in the solar wind is also shown in Fig.~\ref{fig:variance_updated}. For clarity, the solar wind case is shifted, so that $n = 9$ appears as $n = 0$ in the figure. We find that $\sigma_n^2/2 \propto n$ from $n \approx 13$ to $n \approx 19$, with a similar intermittency parameter as for $\epsilon_n/\epsilon_0$ in the weak guide field MHD simulations.

The above results are remarkably consistent with the log-normal model, but do not rule out the log-Poisson model. We note that a comparison is complicated by the fact that the log-Poisson distribution is discrete while the measured PDFs are manifestly continuous. Even in the continuum limit, we are unable to find better fits with the log-Poisson distribution than those with the log-normal distribution, despite having three free parameters. We note, however, that the log-Poisson model only requires that $P(\epsilon_n/\epsilon_0)$ is a convolution of the log-Poisson distribution \citep{dubrulle_1994}; hence, performing log-Poisson fits directly to the PDFs is insufficient to test the log-Poisson model.

\begin{figure}
\includegraphics[width=8cm]{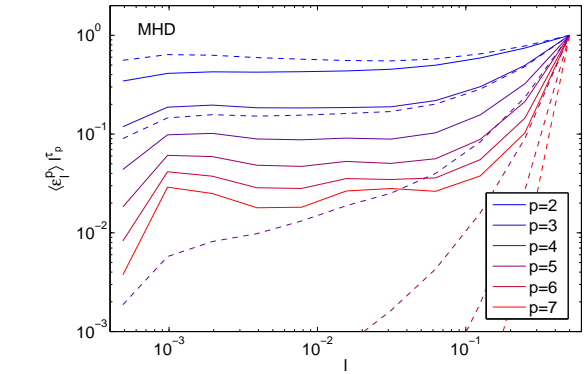}
   \centering
   \caption{\label{fig:moments} Moments of the coarse-grained energy dissipation rate, $\langle \epsilon_l^p \rangle$, versus scale $l$, for $p \in \{2,3,4,5,6,7\}$, compensated by the log-normal prediction (with $\mu = 0.43$; dashed lines) and by the log-Poisson prediction (with $\beta = 0.5$, $C = 2$; solid lines). The normalization is arbitrary.}
 \end{figure}

To determine whether the underlying dynamics are better described by the log-normal model or log-Poisson model, we next consider the scaling of the moments of the coarse-grained energy dissipation rate. In both models, the $p$th moment of the coarse-grained energy dissipation rate is predicted to decrease with increasing scale $l$ as $\langle \epsilon_l^p \rangle \sim l^{-\tau_p}$. The scaling exponents in the log-normal model are given by
\begin{align}
\tau_p^{(LN)} &= \frac{1}{2} m p (p-1), \label{moments:ln}
\end{align}
while the scaling exponents in the log-Poisson model are
\begin{align}
\tau_p^{(LP)} &= C ( 1 - \beta ) p - C ( 1 - \beta^p), \label{moments:lp}
\end{align}
where $C$ is the co-dimension of the most intermittent structure and $\beta$ is the efficiency of energy transfer. In Fig.~\ref{fig:moments}, we show $\langle \epsilon_l^p \rangle$ versus $l$ for $p \in \{2,3,4,5,6,7\}$ in the RMHD simulation with $Re = 9000$, compensated by the predicted scalings for each model. In this case, we perform local averages across squares orthogonal to the guide field\footnote{Cube averages give similar results to square averages.}; the scalings are less robust for line averages and lower resolution simulations. We find that, overall, the log-normal model with $m = 0.43$ can reasonably describe the scalings only up to $p \sim 4$. On the other hand, the scalings can be robustly fit up to $p \sim 7$ by the log-Poisson model. In particular, we find that an excellent fit for this case is given by $\beta = 0.5$ and $C = 2$, implying filamentary structures, in contrast to $C = 1$ proposed by previous models \citep{muller_biskamp_2000}.

\begin{figure*}
\includegraphics[width=8cm]{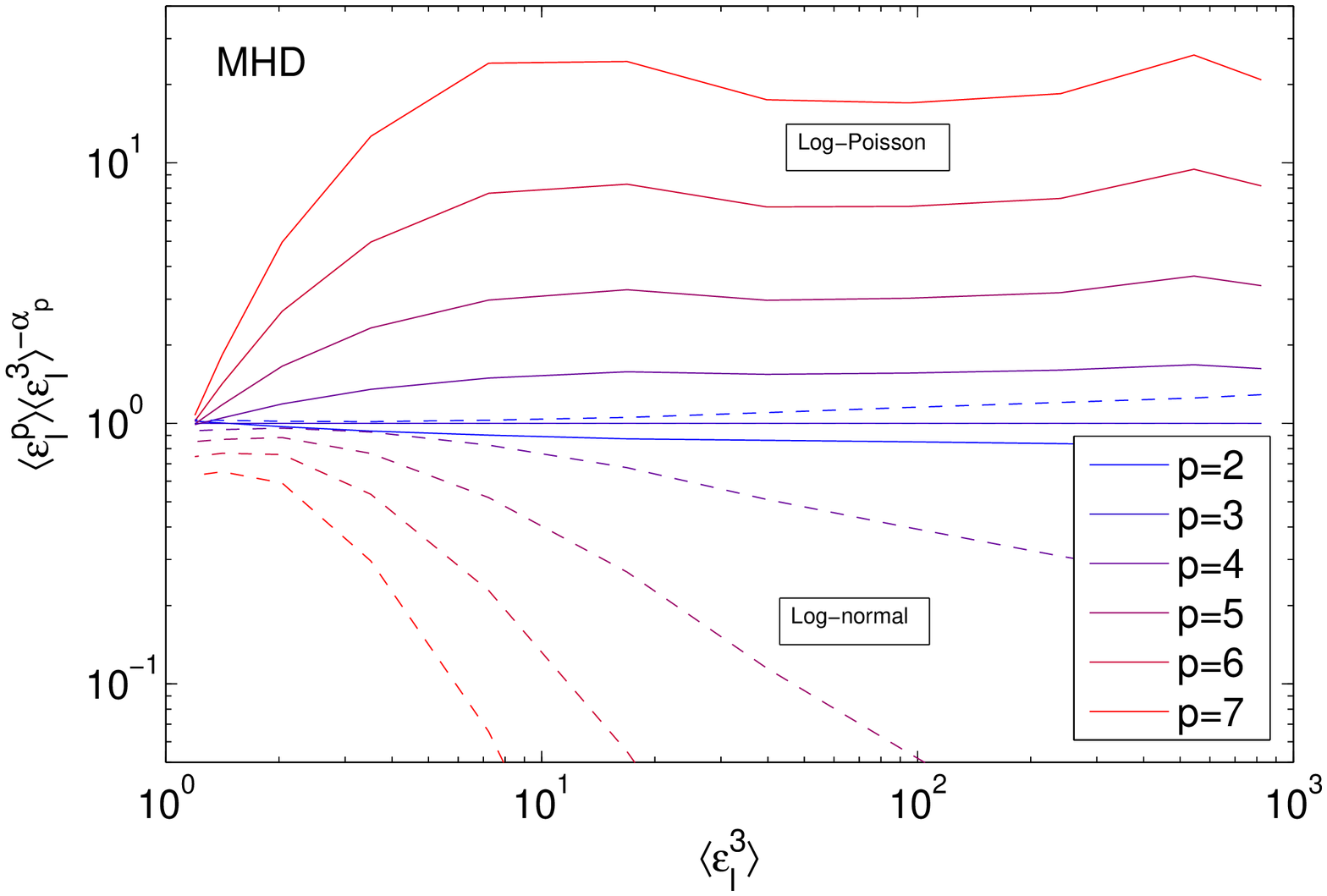}
\includegraphics[width=8cm]{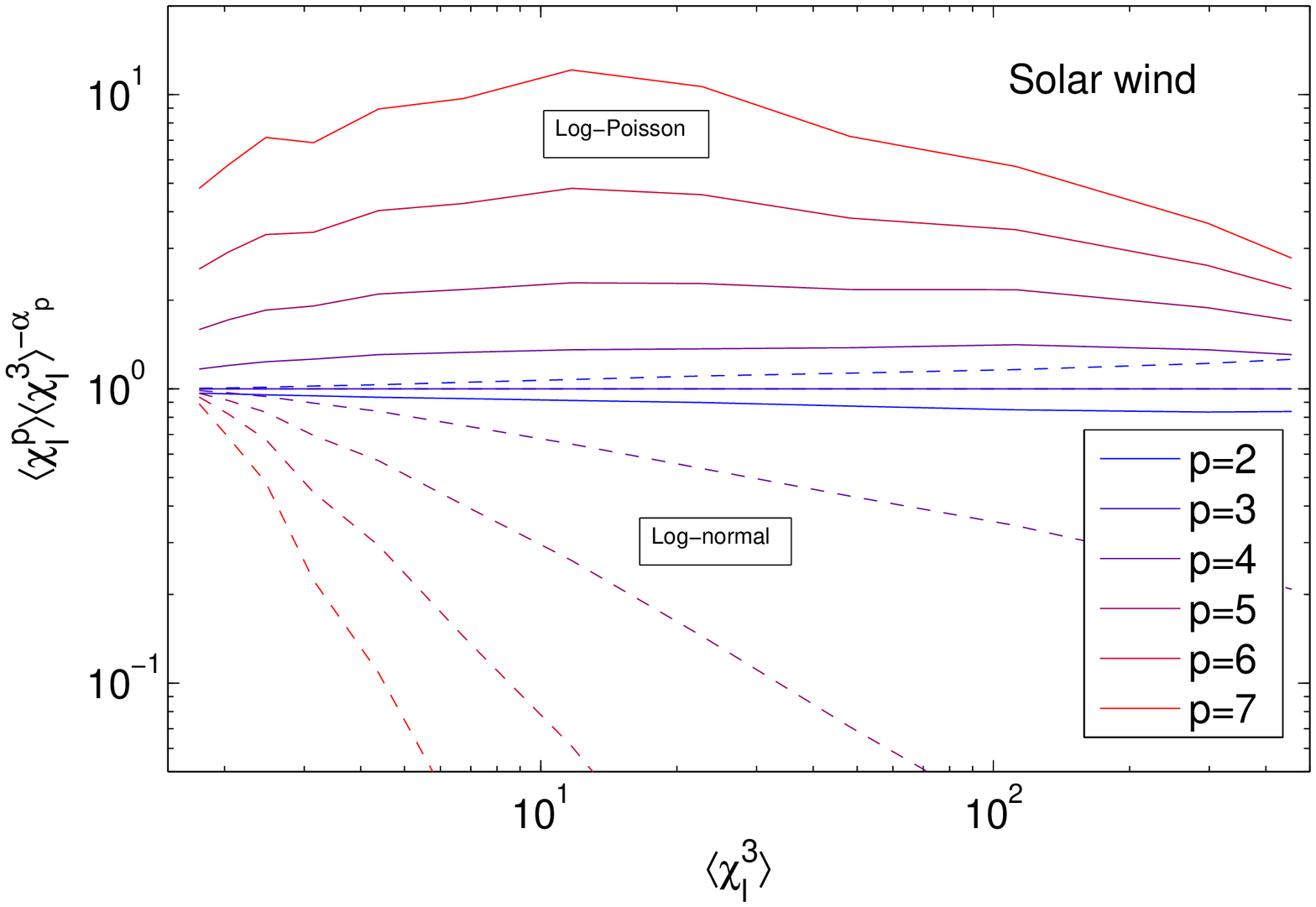}
   \centering
   \caption{\label{fig:ess} Left panel: The $p$th-order moment of the coarse-grained energy dissipation rate, $\langle \epsilon_l^p \rangle$, versus the third-order moment $\langle \epsilon_l^3 \rangle$, for $p \in \{2,3,4,5,6,7\}$, compensated by expected power-law scalings for the log-normal model (dashed lines) and for the log-Poisson model (with $\beta = 0.5$; solid lines). Right panel: Same for solar wind, using moments of surrogate dissipation, $\langle \chi_l^p \rangle$.}
\end{figure*}

We next apply extended self-similarity to get a more robust scaling \citep{benzi_etal_1993}. In this case, we consider the scaling $\langle \epsilon_l^p \rangle \sim \langle \epsilon_l^3 \rangle^{\alpha_p}$ with $\alpha_p = \tau_p/\tau_3$. The predicted scaling exponents $\alpha_p$ are independent of the parameters $m$ for the log-normal model and $C$ for the log-Poisson model. We show $\langle \epsilon_l^p \rangle$ versus $\langle \epsilon_l^3 \rangle$, compensated by the scalings predicted by both models, in the first panel of Fig.~\ref{fig:ess}. In this case, it is clear that the log-Poisson model with $\beta = 0.5$ better describes the scalings than the log-normal model; similar results are obtained for all of our numerical simulations and for different shapes of local averaging regions. Reasonable fits by the log-Poisson model can be obtained for a relatively broad range of $\beta$, including $\beta = 1/3$. For comparison, we show a similar plot for moments of the surrogate dissipation in the solar wind, $\langle \chi_l^p \rangle$, in the second panel of Fig.~\ref{fig:ess}. The results from the solar wind are remarkably similar to the numerical simulations, with the log-Poisson model with $\beta = 0.5$ fitting the observations better than the log-normal model.

\section{Conclusions}

In this work, we investigated the statistics of the coarse-grained energy dissipation rate in two independent systems: numerical simulations of MHD turbulence and the solar wind, using a surrogate quantity in the latter case. We compared the PDFs and moments of the coarse-grained dissipation rate to predictions from two random cascade models: the log-normal model and the log-Poisson model. For both systems, we found that the log-normal model robustly describes the bulk of the PDFs and their scale dependence, with the intermittency parameter being the sole free parameter. This provokes us to consider it as a serious and useful model of intermittency in the Alfv\'{e}nic energy cascade. On the other hand, we found that the log-Poisson model better describes the scalings of the higher-order moments of the coarse-grained dissipation rate, although it has an additional free parameter and does not directly fit the PDF. It is unclear to what extent these high-order moments can be trusted \citep{dudokdewit_2004}; extended self-similarity is required to obtain scalings that are robust in all of the simulations.

This Letter focused on intermittency in the regime of Alfv\'{e}nic turbulence, which is occurs at scales larger than the characteristic microscales (e.g., ion gyroscale or ion skin depth). Random cascade models take the inertial-range energy cascade rate as fundamental, and should therefore be insensitive to the mechanisms of energy dissipation, as long as the dissipation rate is properly measured. However, it is not clear that $j^2$ is a reasonable proxy for dissipation in the solar wind, although the results are remarkably consistent with our MHD simulations. Log-normal distributions were previously measured in the solar wind for magnetic field fluctuations \citep{burlaga_2001, bruno_etal_2001}, rotational discontinuities \citep{zhdankin_etal_2012b, chen_etal_2015}, and proxies for energy cascade rate \citep{sorriso-valvo_etal_2015}, all of which may be indirect surrogates for energy dissipation. We leave it to future work to devise better methods for inferring energy dissipation from solar wind measurements.

\section*{Acknowledgements}

The authors would like to thank Jean Carlos Perez and Joanne Mason for providing data cubes for several of the numerical simulations analyzed in this paper. This research was supported by the NSF Center for Magnetic Self-Organization in Laboratory and Astrophysical Plasmas at the University of Wisconsin-Madison. SB is also supported by the Space Science Institute, by the NASA grant NNX11AE12G, and by the National Science Foundation under the grant NSF AGS-1261659. CHKC is supported by an Imperial College Junior Research fellowship.





\bibliographystyle{mnras}
\bibliography{refs_all} 


\bsp	
\label{lastpage}
\end{document}